\documentclass[universe,article,accept,moreauthors]{Definitions/mdpi}
\firstpage{1} 
\pubvolume{1}
\issuenum{1}
\articlenumber{0}
\pubyear{2023}
\copyrightyear{2023}
\datereceived{ } 
\daterevised{ } 
\dateaccepted{ } 
\datepublished{ } 
\hreflink{https://doi.org/} 

\Title{Using Machine Learning for Lunar Mineralogy-I: Hyperspectral Imaging of Volcanic Samples}

\TitleCitation{Title}

\Author{Fatemeh Fazel Hesar $^{1,4}$\orcidB{},
Mojtaba Raouf  $^{2,3,4*}$\orcidA{} ,
Peyman Soltani $^5$,
Bernard Foing $^{2,4}$,
Michiel J.A. de Dood $^5$,
Fons J. Verbeek $^1$,
Esther Cheng $^4$,
Chenming Zhou $^4$
}

\AuthorNames{Fatemeh Fazel Hesar,Mojtaba Raouf et al. }

\AuthorCitation{Fazel Hesar, F.; Raouf, M.; et al.}

\address{
$^{1}$ \quad Leiden Institute of Advanced Computer Science, P.O. Box 9504, 2300 RA, Leiden, The Netherlands \\
$^{2}$ \quad Leiden Observatory, Leiden University, P.O. Box 9513, 2300 RA, Leiden, The Netherlands \\
$^{3}$ \quad School of Astronomy, Institute for Research in Fundamental Sciences (IPM), 19395-5746, Tehran, Iran\\
$^{4}$ \quad ILEWG LUNEX-EuroMoonMars Earth-Space, ESTEC European Space Agency, Keplerlaan 1, 2201 AZ, Noordwijk, The Netherlands\\
$^{5}$ \quad  Huygens-Kamerlingh Onnes Laboratory, Leiden University, Niels Bohnweg 2, 2333 CA, Leiden, The Netherlands}

\corres{Correspondence: mojtaba.raouf@gmail.com, raouf@strw.leidenuniv.nl}

\abstract{
This study examines the mineral composition of volcanic samples similar to lunar materials, focusing on olivine and pyroxene. Using hyperspectral imaging (HSI) from 400 to 1000 nm, we created data cubes to analyze the reflectance characteristics of samples from samples from Vulcano, a volcanically active island in the Aeolian Archipelago, north of Sicily, Italy, categorizing them into nine regions of interest (ROIs) and analyzing spectral data for each.
We applied various unsupervised clustering algorithms, including K-Means, Hierarchical Clustering, Gaussian Mixture Models (GMM), and Spectral Clustering, to classify the spectral profiles. Principal Component Analysis (PCA) revealed distinct spectral signatures associated with specific minerals, facilitating precise identification. Clustering performance varied by region, with K-Means achieving the highest silhouette score of 0.47, whereas GMM performed poorly with a score of only 0.25. Non-negative Matrix Factorization (NMF) aided in identifying similarities among clusters across different methods and reference spectra for olivine and pyroxene.
Hierarchical clustering emerged as the most reliable technique, achieving a 94\% similarity with the olivine spectrum in one sample, whereas GMM exhibited notable variability. Overall, the analysis indicated that both Hierarchical and K-Means methods yielded lower errors in total measurements, with K-Means demonstrating superior performance in estimated dispersion and clustering. Additionally, GMM showed a higher root mean square error (RMSE) compared to the other models. The RMSE analysis confirmed K-Means as the most consistent algorithm across all samples, suggesting a predominance of olivine in the Vulcano region relative to pyroxene. This predominance is likely linked to historical formation conditions similar to volcanic processes on the Moon, where olivine-rich compositions are common in ancient lava flows and impact melt rocks. These findings provide a deeper context for mineral distribution and formation processes in volcanic landscapes.
}


\begin{document}

\section{Introduction} \label{sec:introduction}

The study of mineral spectra, particularly pyroxene and olivine, is important for understanding the geological composition of celestial bodies like the Moon \cite{taylor2001}. This understanding not only characterizes their history and evolution but also enhances our knowledge of their potential resources \cite{Jaumann2012}. For example, the lunar surface predominantly comprises basaltic rocks rich in ferromagnetic minerals such as pyroxene and olivine, which relate to the Moon’s volcanic activity and mantle composition\cite{Moriarty2021}. Olivine-rich compositions are common in ancient lava flows and impact melt rocks \cite{McSween2006}. This is especially true in lunar geology, where olivine plays a crucial role in the Moon's crust, particularly in areas influenced by ancient volcanic activity and meteorite impacts. These findings enhance our understanding of mineral distribution and formation processes in volcanic landscapes.

The interpretation of lunar data heavily relies on correlations with samples returned from the Moon and analyzed in terrestrial laboratories \cite{Pieters1986}. Hyperspectral imaging has emerged as a powerful tool for analyzing the mineralogical composition of geological samples. Unlike traditional multispectral imaging, which captures data in a limited number of broad spectral bands, hyperspectral imaging acquires data across hundreds of narrow, contiguous spectral bands. This high spectral resolution enables the detection of subtle variations in mineral composition, making it particularly useful for studying complex geological materials such as pyroxene and olivine \cite{VanDerMeer2012}. Hyperspectral data, when combined with advanced machine learning techniques, can significantly enhance the accuracy and efficiency of mineral identification and classification.
These terrestrial analogs allow for a better understanding of the spectral characteristics of lunar materials under varying conditions. Optical instruments measuring reflected visible to near-infrared (VNIR) radiation are sensitive to surface mineralogy due to distinctive absorption features in this wavelength range. Pyroxenes, for instance, display diagnostic absorptions at 1 and 2 µm, while olivine exhibits a combined absorption near 1 µm \cite{Adams1974}. These absorption features shift predictably based on mineral composition \cite{Cloutis1991}.

Recent advances in machine learning have revolutionized the analysis of hyperspectral data, enabling more accurate and efficient mineral identification. Machine learning algorithms, particularly unsupervised clustering methods, have proven effective in identifying patterns and structures within complex spectral datasets. Techniques such as K-means clustering, hierarchical clustering, and Gaussian mixture models are widely used to group similar spectral signatures, facilitating the identification of distinct mineral phases \cite{Plaza2011}. These methods are particularly valuable for analyzing lunar samples, where the spectral signatures of minerals can be subtle and overlapping. The primary objective of this research is to enhance the understanding of lunar geology through detailed mineral analysis, with a particular focus on olivine and pyroxene. This study focuses on a dataset comprising pyroxene and olivine samples collected from Vulcano, Italy, part of the Aeolian Islands, as terrestrial analogs for lunar materials~\cite{Spampinato2011}. Such volcanic regions, provide valuable insights for optimizing hyperspectral measurements of lunar rocks due to their compositional similarities with basaltic lunar surfaces.

This paper is structured as follows:  Section~\ref{sec:Methodology} details the methodology employed in our analysis; 
Section~\ref{sec:Result} presents the results of our findings; and Section~\ref{sec:discussion},\ref{sec:conc} provides a discussion and conclusion of the implications of these results.

\begin{figure}[ht]
\centering
\includegraphics[width=0.8\linewidth]{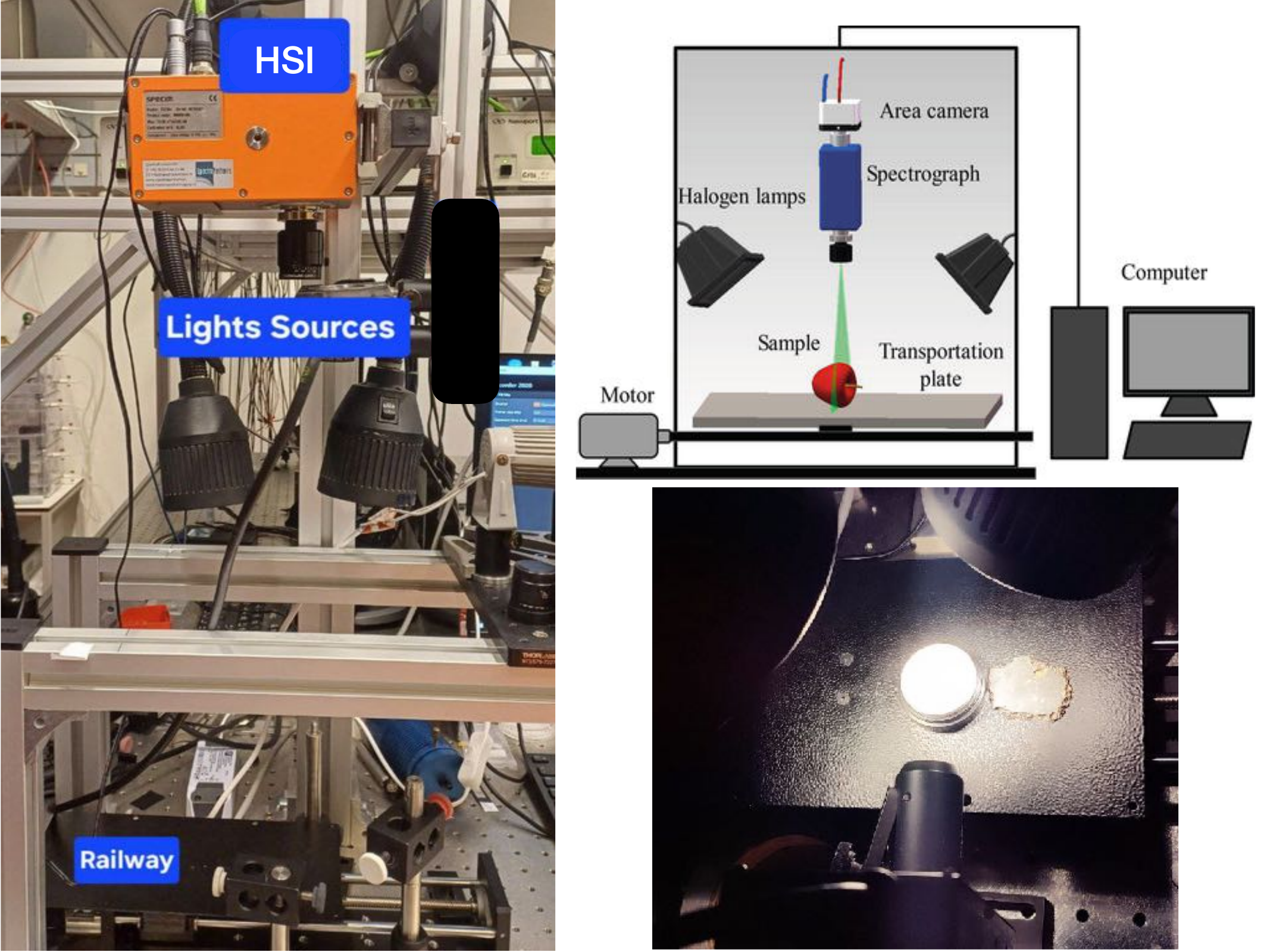}
\caption{Overview of the hyperspectral imaging setup. (Left) The Hyperspectral Imaging System (HSI) with labeled light sources (halogen lamps) and components. (Top-right) Diagram of the experimental arrangement, clearly annotating the area camera, spectrograph, halogen lamps, transportation plate with sample, and connected computer for data acquisition. (Bottom-right) Close-up view of a sample under illumination. The white reference was obtained using a Teflon whiteboard/Spectralon reference with 99\% reflectance.}
\label{fig:Setup_all}
\end{figure}

\begin{figure}[ht]
\centering
\includegraphics[width=0.44\linewidth]{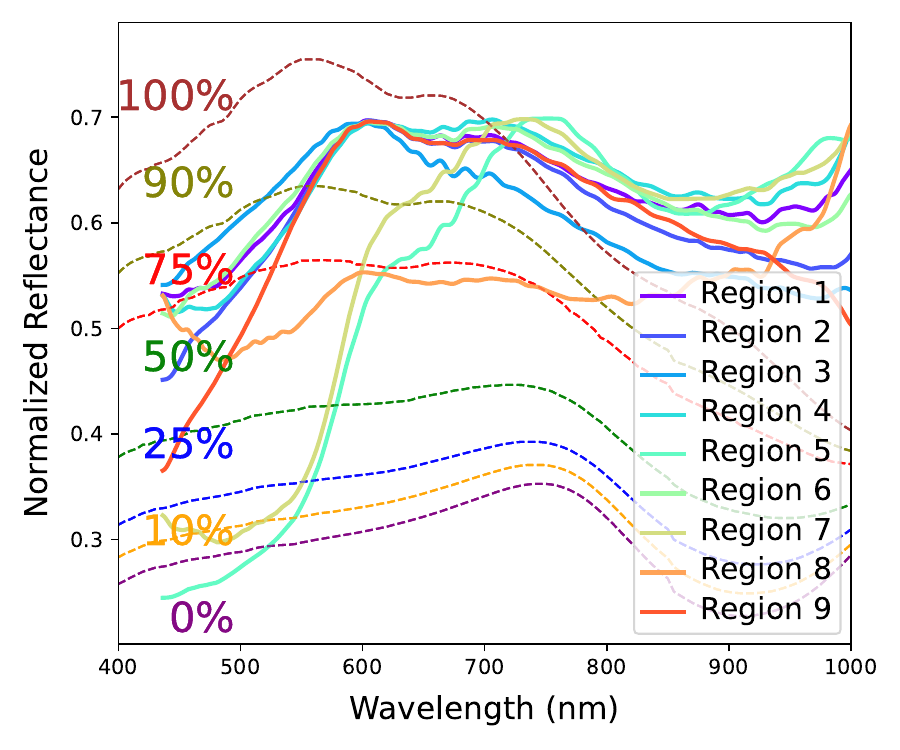}
\includegraphics[width=0.53\linewidth]{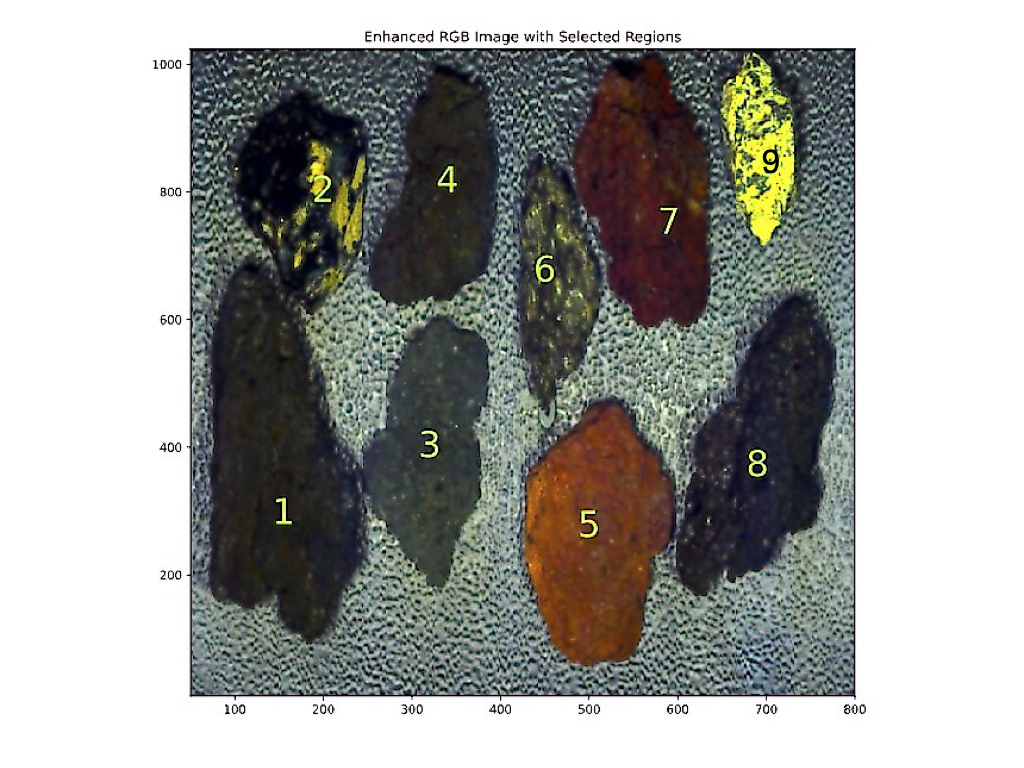}
\caption{(Right) The image displays the RGB reflectance spectra (combining 450, 550, and 650 nm) for nine distinct regions from Vulcano, a volcanically active island in the Aeolian Archipelago, north of Sicily, Italy, covering the wavelength range of 400-1000 nm. (Left) The average spectral responses of these regions reveal variations in reflectance attributed to differences in mineral composition, particularly the amounts of olivine and pyroxene present. The percentage reported for each dashed-color spectrum indicates the proportion of olivine in the olivine-pyroxene composition template from \cite{Mandon2022}. The solid-color lines represent the median spectrum for each region, labeled with specific numbers in the right panel.}
\label{fig:Similarity_regions_Volcano}
\end{figure}

\begin{figure}[ht]
\centering
\includegraphics[width=1\linewidth]{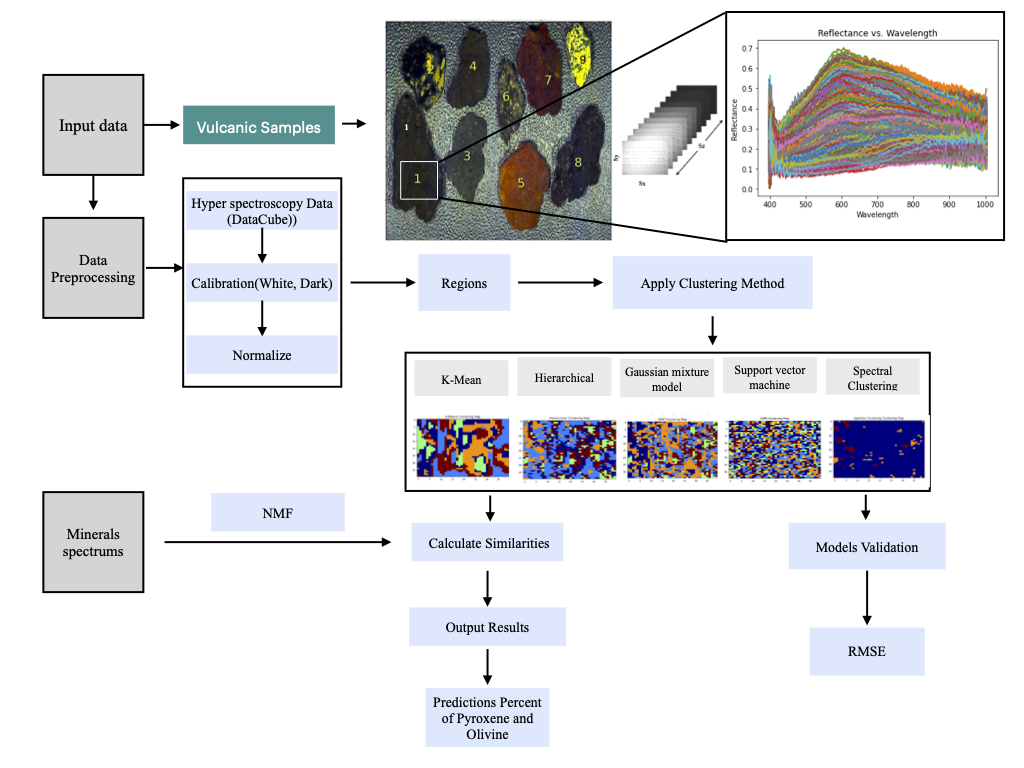}
\caption{Hyperspectral data analysis pipeline for volcanic samples. This figure illustrates the systematic approach used to process and analyze hyperspectral images, including calibration, normalization, and the application of various clustering algorithms. The pipeline ultimately predicts the mineral composition, specifically focusing on pyroxene and olivine, enhancing our understanding of volcanic material characteristics.}
\label{fig:Pipline}
\end{figure}

\begin{figure}[htbp]
    \centering
    \includegraphics[width=0.8\textwidth]{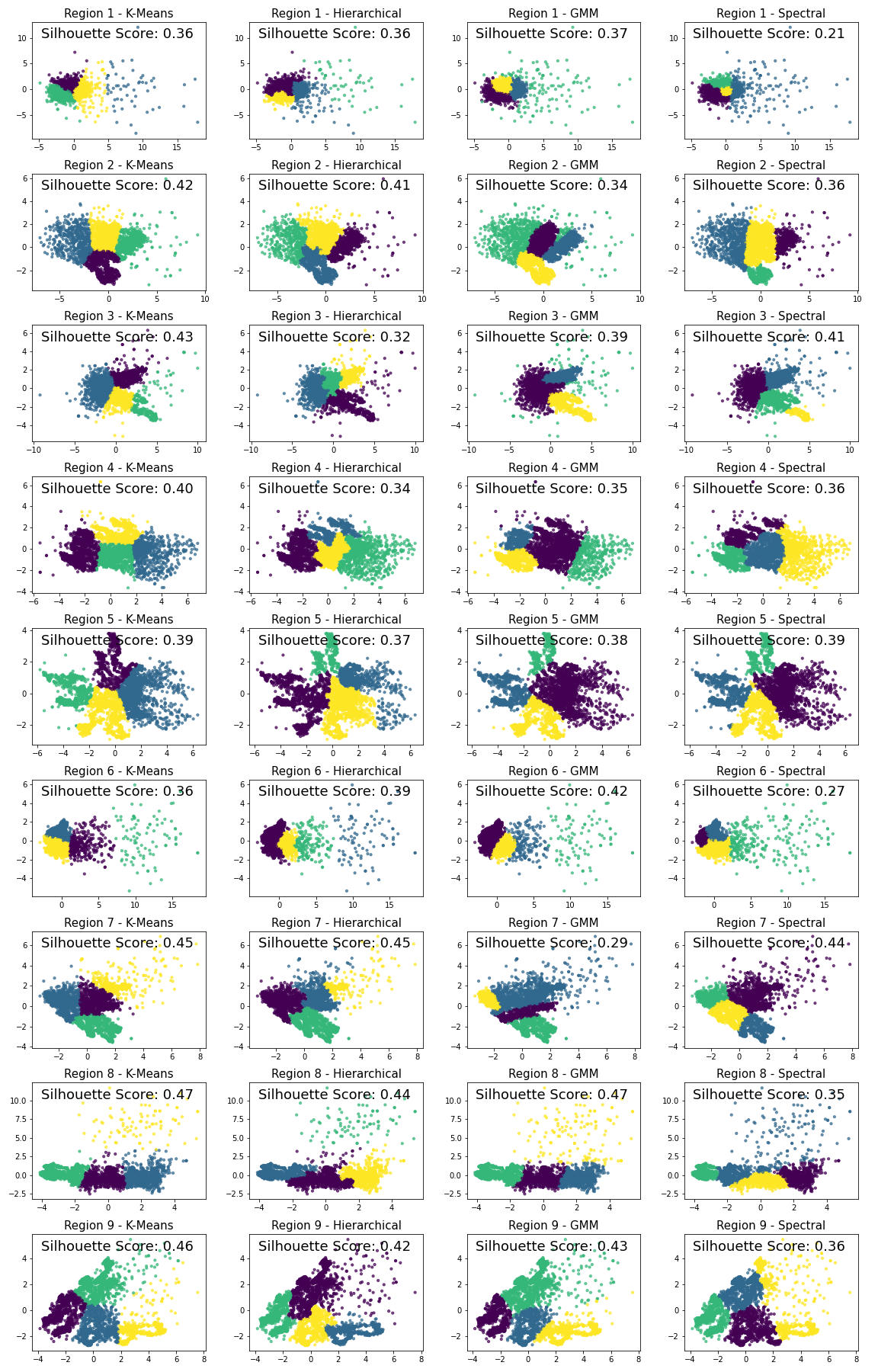}
    \caption{ The visualization presents the PCA-based clustering results for volcanic hyperspectral data across nine regions (Regions 1–9), evaluated using four methods: KMeans, Hierarchical, Gaussian Mixture Model (GMM), and Spectral Clustering, with $n=4$ clusters. Each panel corresponds to a specific region and clustering technique, accompanied by Silhouette Scores to assess cluster quality (with higher scores indicating better-defined clusters). Notable findings include strong clustering performance in Region 8 (KMeans: 0.47) and Region 8 (GMM: 0.47), while weaker performance is observed in Region 1 (Spectral: 0.21) and Region 6 (Spectral: 0.27).}
    \label{fig:PCA}
\end{figure}

\begin{figure}[ht]
\centering
\includegraphics[width=1.1\linewidth]{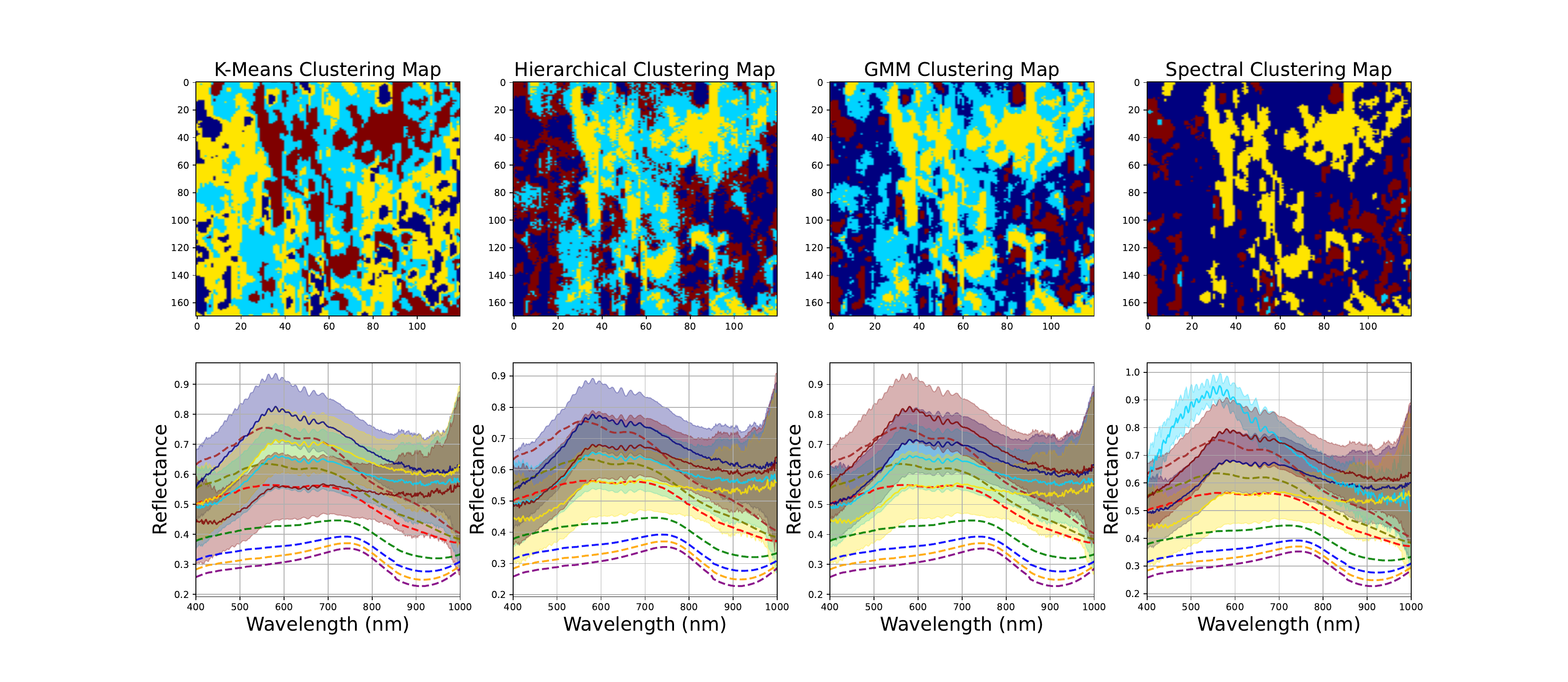}
\caption{(Top) Maps illustrating the effectiveness of four clustering algorithms—K-Means, Hierarchical, Gaussian Mixture Model (GMM), and Spectral Clustering—in grouping mineral compositions into n=4 clusters within a selected 140x190 pixel area of Volcano Region 1.
(Bottom) Comparative visualization of olivine and pyroxene compositions (dashed lines, see Figure \ref{fig:Similarity_regions_Volcano}) across clustering models, shown by solid lines with a $\sigma$ error shaded scatters.}
\label{fig:ML_clustering_model_Map_150_210_RBF}
\end{figure}

\begin{figure}[htbp]
    \centering
    \includegraphics[width=0.99\textwidth]{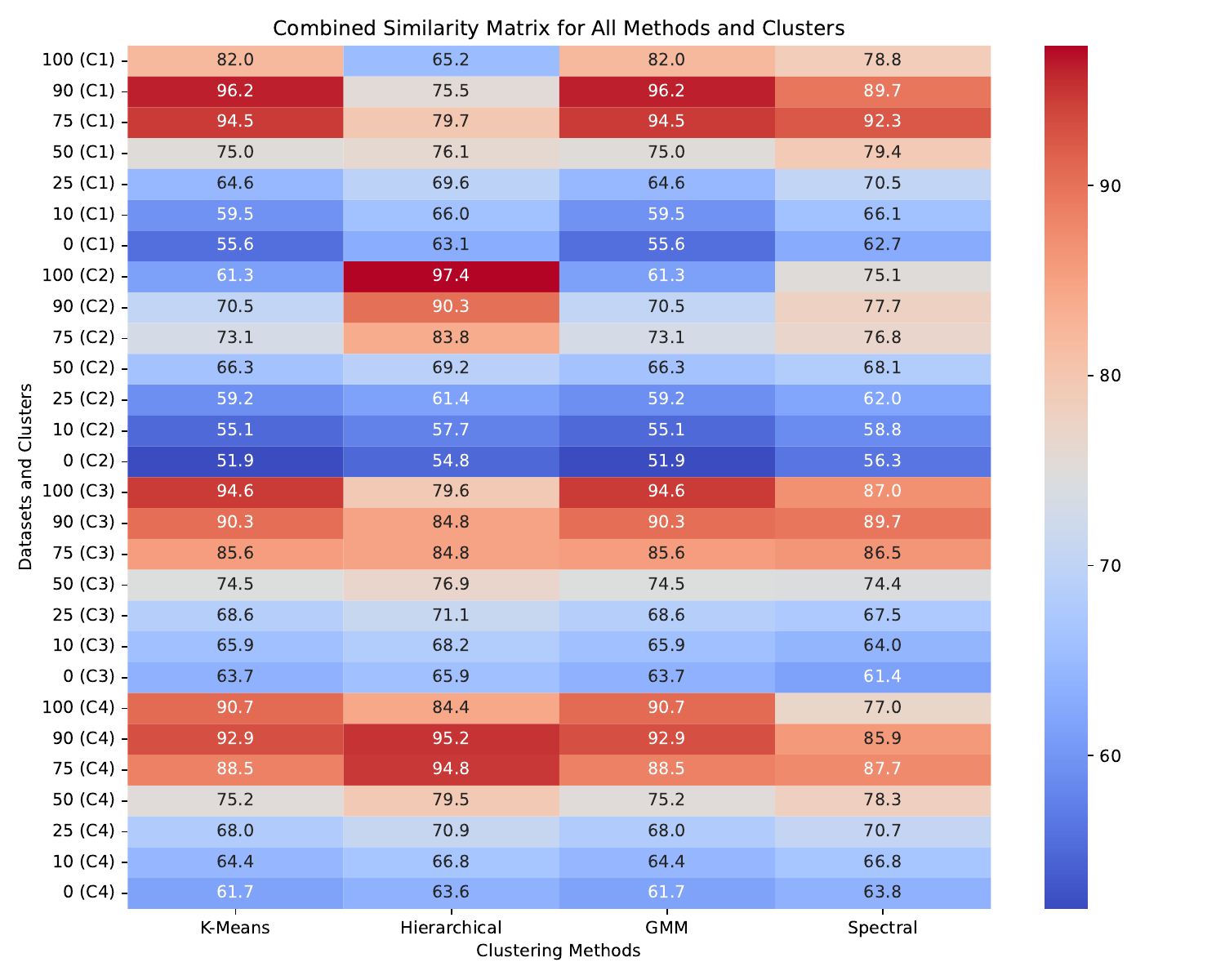}
    \caption{Comparative analysis of mineral composition (the percentage of olivine (100\%-Pyroxene)) similarity across clustering models using NMF. This heatmap illustrates the effectiveness of four clustering algorithms—K-Means, Hierarchical, Gaussian Mixture Model (GMM), and Spectral Clustering—in grouping mineral compositions, specifically Olivine (O) and Pyroxene (P), across different clusters (C1–C4). Each cell represents the similarity percentage for specific mineral mixtures, with color gradients indicating the strength of mineral presence from high (red) to low (blue). The results highlight the unique clustering tendencies of each algorithm.}
    \label{fig:combined_similarity_heatmap}
\end{figure}

\section{Data and Methodology}\label{sec:Methodology} 

This study utilizes an advanced hyperspectral imaging system to capture detailed spectral and spatial data from volcanic samples. The analysis is based on hyperspectral image cube data obtained from volcanic samples collected in Vulcano, a volcanically active island in the Aeolian Archipelago, north of Sicily, Italy, primarily consisting of pyroxenes and olivines. This dataset allows for in-depth examination of the significance of various wavelengths in mineral identification, contributing to our understanding of mineralogy. Moreover, the findings highlight similarities between terrestrial and lunar mineralogy, enhancing our knowledge of these geological materials.

\subsection{Hyperspectral imaging Setup} \label{sec:hyperspectral_setup}
This system includes a grating-based hyperspectral camera with a CMOS sensor (Specim FX10), halogen lamps, and a 20 cm wide lab scanner. The hyperspectral imager functions as a line-scan camera with 1024 pixels, generating 224 spectral images across a wavelength range of 400-1000 nm, with  a spectral full width at half maximum (FWHM) of 5.5 nm. Functioning as a line-scan camera, it captures 1024 spatial pixels per hyperspectral frame or scanned cross-track line.
To ensure accuracy and minimize imaging artifacts, the entire setup was operated in a controlled dark room environment, with samples carefully positioned on the lab scanner platform for imaging. The hyperspectral imaging system employed in this study, based on the Specim FX10, is designed to provide detailed mineral composition of volcanic samples from Vulcano, a volcanically active island in the Aeolian Archipelago, north of Sicily, Italy. The hyperspectral imaging system employed in this study enables the identification of subtle spectral features associated with various minerals.
As illustrated in Figure \ref{fig:Setup_all}, the central component of this setup is the grating-based hyperspectral imager (Specim FX10). This capability allows for identifying subtle spectral features associated with various minerals. The system also includes essential components such as halogen lamps for consistent illumination, a motorized transportation plate for precise sample positioning, and a 20 cm wide lab scanner for facilitating the line-scan imaging process. The experimental arrangement in Figure \ref{fig:Setup_all} highlights the meticulous organization of the system’s components. The halogen lamps are strategically positioned to ensure optimal lighting, while the transportation plate facilitates the accurate movement of samples under the imaging apparatus. Additionally, preprocessing steps, including dark current correction and radiometric calibration using a white reference panel, enhance the accuracy of the reflectance data collected. This structured approach enables the extraction of reflectance values that are used for analyzing the mineralogical characteristics of the samples.

\subsection{Data cube Creation with Hyperspectral Imaging (HSI)} \label{sec:cube_data_creation}
Hyperspectral imaging transcends traditional imaging by capturing far more detailed spectral information. Instead of a simple RGB (red, green, blue) value per pixel, each spatial location (x, y) within the image is assigned a spectrum of reflectance or radiance values across a wide range of wavelengths. This expansion from a two-dimensional array of pixels to a three-dimensional data cube is what allows for the creation of rich, detailed datasets. Each 'voxel' (volume pixel) in this cube holds a complete spectral signature, providing a far more comprehensive representation of the material at that specific location. The spectral dimension ($\lambda$) of the cube represents the reflection of light measured at various wavelengths, often spanning the visible and near-infrared regions of the electromagnetic spectrum. 448 contiguous wavelength bands are recorded by the camera, which is binned to 224 effective wavelength bands, creating a detailed spectral fingerprint for each location.
This detailed spectral information is for distinguishing materials based on their diagnostic spectral features, a technique particularly valuable in fields like geology and mineralogy.

Figure \ref{fig:Similarity_regions_Volcano} presents a hyperspectral image (400-1000 nm) of volcanic samples from Vulcano, a volcanically active island in the Aeolian Archipelago, north of Sicily, Italy, illustrating the effectiveness of hyperspectral imaging in revealing detailed mineral compositions.
The RGB color images utilize typical ranges of R: 650 nm, G: 550 nm, and B: 450 nm to approximate true-color visualization, in line with standard practices, highlighting the extracted regions of interest (ROIs). The extracted regions samples of interest (ROIs) are highlighted by green numbers overlaid on the image, with detailed spectral analysis presented in the left panel. Nine samples, marked by green numbers, were selected for an in-depth spectral analysis of pyroxene and olivine compositions. The left panel of the figure shows the median spectra for the selected regions depicted in the right panel. Notably, at least 80\% of the volcanic samples align with the median spectra for pyroxene and olivine, suggesting that these minerals should ideally be present in all samples, except for Regions 5 and 7, which require further analysis. The database template for pyroxene and olivine, as reported by \cite{Mandon2022}, is overlaid in different colors to represent various ranges of olivine-pyroxene compositions. The percentage of olivine (100\%-Pyroxene) composition is indicated with the corresponding color in the figure.

\subsection{Significance of spectral data} \label{sec:wavelength_significance}

The reflective properties of different minerals and rocks on the Moon, particularly at specific optical wavelengths, enable us to distinguish between them, building on prior studies of pyroxene and olivine spectral features \cite{Adams1974,Cloutis1991}. This work applies these features to hyperspectral data (400--1000~nm) from Vulcano samples, using machine learning to classify mineral compositions across specific ROIs.

This capability is for lunar studies, as understanding the composition of the Moon's surface can provide its geological history and processes. For instance, the ultraviolet and blue ranges (400-500 nm) are particularly useful for identifying minerals such as pyroxene and olivine, which exhibit diagnostic absorption features in these wavelengths. These minerals are often found in the lunar highlands and are significant indicators of the Moon's volcanic history. Additionally, the violet range (400-420 nm) aids in studying the presence of impact glasses formed during meteoroid impacts, which are formed when meteoroids strike the lunar surface at high velocities, melting and fusing the surrounding materials.

The green wavelength range (500-550 nm) is significant for detecting iron-bearing minerals like basalt, which helps scientists identify regions rich in basaltic material predominantly found in the lunar maria. These dark, flat plains were formed by ancient volcanic activity, providing evidence of the Moon's geological evolution. Furthermore, the yellow range (570-590 nm) is essential for assessing the maturity of lunar regolith, indicating the extent of space weathering and alteration that surface materials have experienced due to continuous bombardment by meteoroids and solar radiation. The orange range (590-630 nm) highlights the presence of iron oxides, such as hematite, which can form due to both meteoroid impacts and solar wind interactions with the lunar surface.

 Hyperspectral imaging and analysis of reflectance spectra at specific wavelengths facilitate the identification of several key minerals on the lunar surface, enhancing understanding of the Moon's geology and its response to meteoroid impacts. Olivine, which is rich in magnesium and iron, is typically detected between 600-700 nm and is commonly found in igneous rocks such as basalt and peridotite. Plagioclase feldspar, also appearing in the 600-700 nm range, is prevalent in many igneous rocks formed from the solidification of molten rock. This mineral exhibits a range of colors and a vitreous to pearly luster, making it an important component of the lunar crust \cite{Pieters1986}.
Pyroxene minerals display distinct absorption features in both the visible and near-infrared ranges, aiding in their identification in reflectance spectra. These minerals are critical in understanding the thermal history of the Moon \cite{Moriarty2021}.

\subsection{Data Acquisition and Preprocessing} 

This study acquired hyperspectral data from volcanic samples in the Vulcano region, Italy, capturing reflectance spectra across 400-1000 nm. The data cubes have a three-dimensional structure of 818 × 1024 × 224, with 224 representing the number of spectral bands. The linear scanner, with a 200 mm width and an exact 200 mm scanning range in the x-direction, captures data where each pixel measures approximately 0.25 mm in the x-dimension (calculated as 200 mm / 818 pixels $=$ 0.24 mm) and $\sim$ 0.2 mm in the y-dimension (calculated as 200 mm / 1024 pixels $=$ 0.195 mm, rounded to 0.2 mm for simplicity).

The pipeline details of the analysis in this study are shown in Figure \ref{fig:Pipline}. We implemented a comprehensive preprocessing pipeline for the hyperspectral cube data, focusing on enhancing data integrity and optimizing clustering performance. This process began with normalizing reflectance values using white and dark reference measurements, effectively addressing illumination variations and sensor noise that can obscure spectral features. Calibration steps, including dark current correction and radiometric calibration, ensured accurate spectral analyses by mitigating systematic errors that could influence subsequent clustering results. Additional techniques, such as outlier detection and removal, were employed to eliminate anomalous data points that could skew results. This meticulous approach ensured high-quality, well-structured hyperspectral data, laying a robust foundation for reliable clustering outcomes and enhancing our ability to accurately interpret the mineralogical characteristics of volcanic materials.
Data acquisition utilized a line scanning technique, with each sample positioned on a motorized slider table. As the table advanced, the camera captured the spectral data in a sequential line-by-line manner, producing an initial hyperspectral image. Calibration minimized noise and ensured the accuracy of reflectance values, achieved through dark and white reference measurements.The dark reference was established by measuring the signal from the detector when no light was present, while 
the white reference was obtained using a Teflon whiteboard/Spectralon white reference with 99\% reflectance (shown in the bottom right panel of Figure \ref{fig:Setup_all}). Subsequently, the calibration process normalized the hyperspectral images using the following formula:
\begin{equation}
    \text{Reflectance} = \frac{(\text{Raw} - \text{Dark})}{(\text{White} - \text{Dark})},
\end{equation}
where \( \text{Raw} \) represents the initial signal that includes both the actual light from the sample and any noise or background signals.

\subsection{Machine Learning: Clustering Methods} \label{sec:Machine Learning: Clustering Methods}
We employed a range of unsupervised machine learning clustering algorithms to analyze hyperspectral data (400--1000 nm) from volcanic samples, aiming to identify mineral compositions, primarily olivine and pyroxene. The procedure involved preprocessing data cubes to ensure quality, followed by applying various clustering methods to group similar spectral signatures. We utilized K-Means for its efficiency in partitioning large datasets, Hierarchical Clustering for its structured approach through dendrograms, Gaussian Mixture Models (GMM) for probabilistic modeling of complex shapes, and Spectral Clustering for detecting non-convex clusters using eigenvalues. To optimize processing time and reduce redundancy in adjacent spectral channels without data loss, we applied dimensionality reduction techniques such as Principal Component Analysis (PCA) and Non-Negative Matrix Factorization (NMF). This comprehensive framework not only preserved key spectral features but also validated clusters against reference spectra, ultimately enhancing our understanding of the mineralogical characteristics present in the samples.

By utilizing multiple clustering approaches, we aimed to extract relevant features from the data, improve analytical accuracy, and gain a deeper the spectral characteristics of the volcanic samples. This multifaceted strategy allows for a comprehensive exploration of the complexities and variabilities inherent in the data. Using diverse models is for revealing different aspects of the dataset, thereby enhancing our understanding of spectral characteristics, as illustrated in the clustering maps. Relying solely on a single method could introduce bias, so employing multiple models validates our findings and ensures consistency in cluster identification across approaches. Given the complexity of hyperspectral data, which often displays non-linear relationships and variable cluster shapes, it is essential to apply different algorithms to capture a broader range of structures.

Among the methods utilized, K-Means clustering stands out for its computational efficiency, particularly suited for the large datasets typical of hyperspectral imaging. This algorithm partitions data points into clusters by minimizing within-cluster variance, effectively delineating distinct divisions within the data. This approach is particularly effective for identifying mineralogical features, with a focus on pyroxenes and olivines.
Additionally, hierarchical clustering is employed to construct a dendrogram that visually represents relationships among the data. This technique computes distances between clusters using various linkage criteria, offering the hierarchical structure of the mineral compositions. The hierarchical clustering map enriches our understanding of how different mineral types interrelate, revealing patterns that may be obscured by other methods. To further enhance our analysis, GMM are utilized. GMMs effectively model the data's probability distribution as a mixture of Gaussian distributions, capturing complex cluster shapes. Lastly, spectral clustering uses the eigenvalues of a similarity matrix, enabling dimensionality reduction while preserving significant data structures. Collectively, these methodologies provide a robust framework for understanding the mineralogical characteristics of volcanic samples.

We employed dimensionality reduction techniques to optimize our data analysis, utilizing PCA to retain significant spectral information while minimizing redundancy and enhancing computational efficiency. Additionally, we explored NMF as a complementary method, which decomposes the dataset into non-negative components that align with the physical interpretation of spectral data. NMF not only aids in reducing dimensionality and managing redundancy but also enhances interpretability by facilitating the identification of specific mineral signatures. Furthermore, it improves the signal-to-noise ratio, allowing for more accurate mineral identification by separating meaningful spectral information from noise. By combining NMF with PCA, we enhance our ability to extract endmember spectra, providing a clearer understanding of the mineralogy in our samples.

\subsubsection{PCA Analysis} \label{sec:PCA Analysis}

Principal Component Analysis (PCA) was applied to the hyperspectral data extracted from various regions of interest within the volcano dataset. The first two principal components (PC1 and PC2) were visualized to assess the underlying structure and variability of the data. Figure \ref{fig:PCA} illustrates the PCA results for each region, providing information on the homogeneity and distribution of the data. Each point in the scatter plot represents a pixel within the region, colored according to its cluster membership as determined by clustering method (K-means, hierarchical, GMM and Spectral), with $n=4$\footnote{The choice of 
$n=4$ clusters is optimal, as both performance metrics and visual assessments indicate superior clustering quality compared to other cluster numbers.} clusters. The spread of the points indicates the diversity of spectral characteristics, while the clustering reveals distinct groups of similar spectral signatures. 
The performance of these clustering methods was evaluated using the Silhouette Score, which measures how similar an object is to its own cluster compared to other clusters. Additionally, Root Mean Squared Error (RMSE) was employed as an error metric to quantify prediction accuracy, aiding in fine-tuning spectral interpretations against instrumental verification.

The first method, K-Means clustering, partitions the data into  distinct clusters by minimizing the variance within each cluster. It is widely used due to its simplicity and efficiency, making it suitable for large datasets. For example, in Region 8, K-Means achieved a Silhouette Score of 0.47, indicating the highest among all methods, suggesting excellent cluster separation.
Next, we applied Hierarchical clustering, which builds a hierarchy of clusters either through a bottom-up (agglomerative) or top-down (divisive) approach. This method provides a dendrogram representation, allowing for a visual exploration of the data's structure and the number of clusters. In Region 7, Hierarchical clustering achieved a Silhouette Score of 0.45, indicating well-separated clusters the highest among other regions. 
We also utilized the GMM, a probabilistic model that assumes all data points are generated from a mixture of several Gaussian distributions with unknown parameters. GMM is particularly advantageous for soft clustering, where data points can belong to multiple clusters with different probabilities, making it more flexible than K-Means. Note that, in Region 8, GMM had a highest Silhouette Score of 0.47 same as K-means indicating similarity in both method clustering.
Finally, we employed Spectral Clustering, which utilizes the eigenvalues of a similarity matrix to reduce dimensionality before applying a clustering algorithm. This method effectively identifies clusters in non-convex shapes, leveraging the relationships between points in the original high-dimensional space. In Region 7, Spectral Clustering achieved a Silhouette Score of 0.44, demonstrating moderate cluster separation.

\subsection{Non-Negative Matrix Factorization for Similarity Analysis}\label{sec:NMF}
To quantify the mineral composition of our volcanic samples, we used Non-Negative Matrix Factorization (NMF) to measure the similarity between the spectral signatures of clusters and reference spectra for olivine and pyroxene from~\cite{Korda2023}. NMF is a technique that breaks down complex data, like our hyperspectral reflectance measurements, into simpler, non-negative components. In this study, it helped us determine how much each cluster resembles pure olivine, pure pyroxene, or a mixture of both. We applied NMF to the median spectra of clusters obtained from K-Means, Hierarchical, GMM, and Spectral Clustering ~\ref{sec:Machine Learning: Clustering Methods} . The output provided similarity percentages (e.g., 90--100\% for olivine in several clusters, see Figure \ref{fig:combined_similarity_heatmap}), which we visualized in a heatmap to compare the performance of each clustering method. This step enhances our ability to draw parallels with lunar samples, where precise mineral identification is essential. By linking unsupervised clustering results to known mineral spectra, NMF supports our goal of understanding the mineralogical makeup of Vulcano samples and their relevance to lunar geology (see Section~\ref{sec:discussion}).
In similarity analysis, NMF uncovers patterns among datasets by representing complex relationships in a more interpretable format. By factorizing the user-item interaction matrix, NMF helps identify similarities between different datasets based on their underlying features. This method is especially useful for non-negative data, such as reflectance measurements in spectral analysis. The resulting factor matrices enable the computation of similarity scores, allowing for effective comparison and clustering of similar items or observations.

\section{Results} \label{sec:Result}
\subsection{Mineral composition similarity} \label{sec:onesample}

The vibrancy of the spectral data, depicted in Figure \ref{fig:ML_clustering_model_Map_150_210_RBF}, elucidates the reflectance properties of the minerals (Olivin-Pyroxene compositions) for the rectangle selected in the first region (size:120x170). The spectra, plotted against wavelength, reveal distinct peaks and troughs indicative of mineral presence. For instance, Olivine exhibits characteristic peaks in the near-infrared range, while Pyroxene displays unique spectral signatures. The black line serves as a baseline for comparison, enhancing the understanding of mineral distribution across volcanic terrains. 
We assessed mineral composition similarity, defined as the percentage match between clustered spectra and reference spectra of olivine and pyroxene \cite{Korda2023}, using NMF (Section \ref{sec:NMF}). We focused on these minerals due to their prevalence in lunar basaltic rocks and Vulcano samples. Spectra were compared to establish robust correlations, as shown in Figure \ref{fig:Similarity_regions_Volcano}. This analysis established robust correlations between the average spectra of the selected region and known minerals, specifically focusing on Pyroxene and Olivine. 
We utilized reference spectra from \cite{Mandon2022}, representing typical lunar-relevant compositions (e.g., forsterite-rich olivine, enstatite-rich pyroxene), acknowledging that these mineral groups vary in Mg-Fe or Ca-Mg-Fe ratios and may transform over time. These spectra detail reflectance properties of olivine and pyroxene mixtures, guiding our similarity analysis.

We compared this existing data with the spectra obtained from each cluster in our models to assess the percentage composition of pyroxene and olivine (represented by dashed color spectrums in the bottom panels of Figure \ref{fig:ML_clustering_model_Map_150_210_RBF}), thereby reporting the similarity proportions (from 1\% to 100\%) using NMF techniques describe in Sec.\ref{sec:NMF}.

The results revealed high similarity scores in certain regions, indicating a significant presence of these mineral. Such findings are critical for interpreting the geological characteristics of the volcano region, as the prevalence of Pyroxene and Olivine is often associated with specific volcanic processes and conditions. Analyzing these spectral features not only aids in mineral identification but also informs geological interpretations, contributing to a broader understanding of lunar geology. Figure \ref{fig:combined_similarity_heatmap} presents a heatmap illustrating the similarity in mineral compositions (Olivine percentage) across the four clustering algorithms. The analysis indicates that all methods perform well with pure Olivine compositions, achieving similarity percentages above 90\% in several clusters. However, when examining mixed compositions of Olivine and Pyroxene, variability in performance becomes evident. K-Means and Hierarchical Clustering showed better adaptability to mixed data compared to GMM, which struggled with lower similarity scores in certain cases.

\subsection{The overall consequence for all regions}

To improve the accuracy of our results, we selected 10 random regions from each of the 9 rocks (regions). For each sample, we identified the best performance and highest probability for olivine-pyroxene composition (maximum similarity) as describe above in Sec. \ref{sec:onesample} . We also recorded the median and standard deviation ($\sigma$) of the error values from 10 iterations of sub-sampling for each region (1-9). This approach allows us to quantify the optimal percentage of olivine-pyroxene composition for each region, incorporating all parts of each area. Specifically, in Region 1, K-Means indicates 100\% olivine with 98.09\% similarity, while in Region 6, it shows 75\% olivine (25\% pyroxene) with 95.56\% similarity, highlighting distinct mineral distributions. Table \ref{tab:summary} provides a comprehensive summary of Olivine-rich percentage across nine regions, analyzed using four clustering methods: Hierarchical, K-Means, Gaussian Mixture Model (GMM), and Spectral Clustering. 
Hierarchical clustering consistently achieved maximum compositions of 99\% in the first two regions, demonstrating its effectiveness in accurately capturing olivine-pyroxene ratios. These ratios were derived by comparing clustered spectra to reference compositions via NMF and averaging over 10 iterations (Table~\ref{tab:summary}).

K-Means also performed admirably, reaching a maximum of 98\% in region 1, although it recorded a 75\% Olivine (25\% Pyroxene) in Region 6 with 96\% similarity similar fining in Spectral method in this region. In contrast, GMM exhibited significant variability, particularly in Region 2, where it reported only 25\% Olivine (i.e 75\% Pyroxene) with 98\% similarity but high $\sigma$ error of $\sim 13$ compare to other methods, highlighting its limitations in handling complex mineral distributions.
Statistical metrics further illustrate the strengths and weaknesses of each method. The median values indicate that Hierarchical and K-Means clustering yielded similar central tendencies for olivine-pyroxene composition, with K-Means often showing slightly lower median values. Notably, K-Means exhibited lower standard deviations ($\sigma$), indicating more consistent results across iterations, especially in Regions 1 and 4. In Region 2, GMM's high $\sigma$ of 12.94 underscores its instability in this context, contrasting with the more robust performances of the other methods.

\subsection{Evaluation of methods}\label{sec:Evaluation of methods}

The primary evaluation metric employed is the Root Mean Squared Error (RMSE), which quantifies the differences between predicted cluster assignments and true labels. The RMSE is calculated as follows:
\begin{equation}
    RMSE = \sqrt{\frac{1}{n} \sum_{i=1}^{n} (y_i - \hat{y}_i)^2},
\end{equation}
where \(y_i\) represents the true label and \(\hat{y}_i\) denotes the predicted label for each instance. The results indicate varying performance among the models across different clusters, as shown in Table \ref{tab:summary}.
K-Means consistently displayed the lowest RMSE values across most regions, indicating its accuracy in predicting olivine-pyroxene compositions. For instance, K-Means recorded an RMSE of 0.43 in Region 5 or 0.38 in Region 9, underscoring its precision. Overall, the results highlight the superiority of Hierarchical clustering and K-Means in estimating olivine-pyroxene compositions, while GMM's variability raises concerns about its reliability. These findings enhance our understanding of volcanic regions' geological characteristics and formation processes.

\begin{figure}[htbp]
    \centering
    \includegraphics[width=0.8\textwidth]{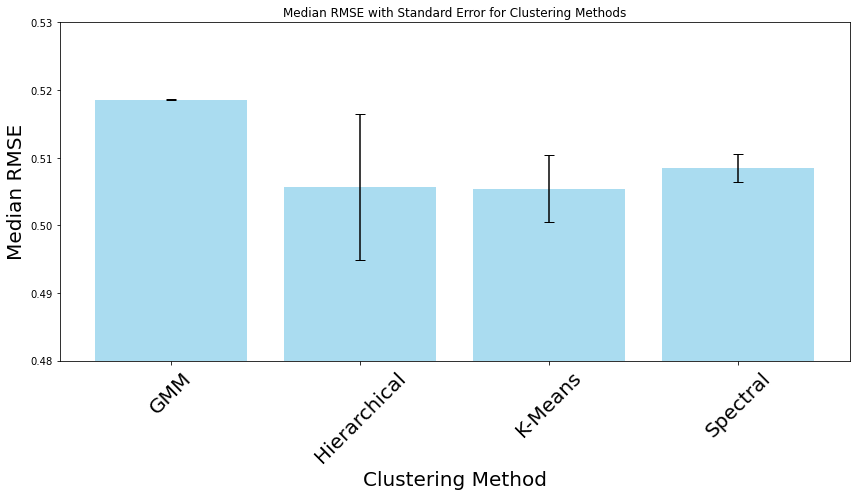}
    \caption{The bar chart presents the median Root Mean Squared Error (RMSE) values for four clustering methods with standard deviation error: GMM, Hierarchical, K-Means, and Spectral. }
    \label{fig:RMSE}
\end{figure}

The bar chart in Figure \ref{fig:RMSE} depicts the RMSE values for the evaluated clustering methods. GMM recorded the highest median RMSE at 0.52, suggesting potential issues with its underlying assumptions about data distribution. Hierarchical and Spectral methods followed closely with a median RMSE of 0.51, while K-Means demonstrated the lowest at 0.50. The standard error bars associated with each method underscore the variability of the RMSE. This research underscores the importance of selecting appropriate clustering algorithms based on the specific characteristics of mineral compositions.

\begin{table}[H]
\centering
\small
\begin{tabular}{cccccccc}
\hline
Region & Method & Olivin(100\%-Pyroxene) & Max Similarity & Median & $\sigma$ & RMSE \\
\hline 
1 & Hierarchical & 100.00 & 99.66 & 97.36 & 3.30 & 0.54 \\
1 & GMM & 100.00 & 97.83 & 93.64 & 1.78 & 0.55 \\
1 & K-Means & 100.00 & 98.09 & 93.79 & 4.12 & 0.52 \\
1 & Spectral & 90.00 & 97.69 & 95.62 & 0.80 & 0.53 \\
\hline
2 & Hierarchical & 100.00 & 99.20 & 89.19 & 5.32 & 0.51 \\
2 & GMM & 25.00 & 97.55 & 79.25 & 12.94 & 0.54 \\
2 & K-Means & 100.00 & 97.08 & 88.94 & 4.08 & 0.53 \\
2 & Spectral & 100.00 & 94.77 & 80.19 & 1.98 & 0.52 \\
\hline
3 & Hierarchical & 100.00 & 85.84 & 79.54 & 3.2 & 0.53 \\
3 & GMM & 100.00 & 87.99 & 78.63 & 1.49 & 0.50 \\
3 & K-Means & 100.00 & 92.71 & 81.35 & 1.53 & 0.48 \\
3 & Spectral & 100.00 & 94.76 & 81.92 & 6.27 & 0.50 \\
\hline
4 & Hierarchical & 100.00 & 96.30 & 95.40 & 0.54 & 0.50 \\
4 & GMM & 90.00 & 97.69 & 89.66 & 4.21 & 0.52 \\
4 & K-Means & 100.00 & 95.58 & 89.84 & 4.05 & 0.53 \\
4 & Spectral & 90.00 & 96.33 & 93.04 & 5.28 & 0.49 \\
\hline
5 & Hierarchical & 100.00 & 84.50 & 72.60 & 3.72 & 0.46 \\
5 & GMM & 100.00 & 79.76 & 76.65 & 2.20 & 0.44 \\
5 & K-Means & 100.00 & 84.36 & 81.10 & 2.04 & 0.43 \\
5 & Spectral & 100.00 & 84.70 & 79.35 & 2.94 & 0.45 \\
\hline
6 & Hierarchical & 100.00 & 97.39 & 76.52 & 2.62 & 0.50 \\
6 & GMM & 100.00 & 93.10 & 77.13 & 2.15 & 0.51 \\
6 & K-Means & 75.00 & 95.56 & 95.56 & 0.00 & 0.54 \\
6 & Spectral & 75.00 & 96.89 & 96.89 & 0.00 & 0.55 \\
\hline
7 & Hierarchical & 100.00 & 94.64 & 83.24 & 3.81 & 0.52 \\
7 & GMM & 100.00 & 94.64 & 87.13 & 2.30 & 0.50 \\
7 & K-Means & 100.00 & 89.85 & 86.57 & 1.40 & 0.49 \\
7 & Spectral & 100.00 & 94.47 & 83.71 & 2.61 & 0.54 \\
\hline
8 & Hierarchical & 90.00 & 93.79 & 88.99 & 2.18 & 0.52 \\
8 & GMM & 90.00 & 96.43 & 91.10 & 1.83 & 0.55 \\
8 & K-Means & 75.00 & 95.68 & 86.09 & 6.78 & 0.57 \\
8 & Spectral & 90.00 & 95.46 & 94.28 & 2.27 & 0.52 \\
\hline
9 & Hierarchical & 100.00 & 90.24 & 75.57 & 2.91 & 0.38 \\
9 & GMM & 100.00 & 90.28 & 72.00 & 2.76 & 0.40 \\
9 & K-Means & 100.00 & 90.28 & 71.45 & 2.85 & 0.38 \\
9 & Spectral & 100.00 & 90.72 & 73.90 & 2.92 & 0.36 \\
\hline
\end{tabular}
\caption{A comparative analysis was conducted on various clustering methods—Hierarchical, K-Means, Gaussian Mixture Model (GMM), and Spectral Clustering—to evaluate olivine-pyroxene composition across nine distinct regions over 10 iterations. The analysis reports key metrics, including maximum similarity, median values, and standard deviation ($\sigma$) for each region and method. Additionally, the root mean square error (RMSE) values are provided in the final column for all methods.}
\label{tab:summary}
\end{table}

\section{Comparison of Lunar Volcanism Minerals}

 Our analysis suggests a predominance of olivine in the Vulcano region, potentially linked to historical conditions akin to lunar volcanic processes. These findings resonate with the conclusions of \cite{Head1976,Corley2018}, who documented the origins of mare lavas on the Moon and noted the significance of olivine-rich compositions, emphasizing that both studies recognize the importance of olivine in volcanic activity.
 \cite{SURKOV2020} provide valuable insights into lunar mineralogy by mapping ilmenite abundance using data from the Chandrayaan-1 M3 mission. Their findings on the relationship between ilmenite and $\text{TiO}_2$ abundances can be related to our study, which focuses on olivine and pyroxene in volcanic samples analogous to lunar materials. Both studies highlight the significance of specific minerals in understanding the Moon's geological history and composition. The methodologies used by \cite{SURKOV2020}  for analyzing absorption features parallel our approach with hyperspectral imaging, emphasizing the importance of spectral data in mineral identification. Furthermore, their conclusion that titanium is not exclusively found in ilmenite suggests broader implications for understanding other titanium-bearing minerals, which may relate to the mineral composition we observe in our Vulcano samples. \cite{Bhatt2019} focus on the quantitative estimation of elemental concentrations on the lunar surface using hyperspectral near-infrared (NIR) imaging. Their study aims to map the abundances of iron (Fe), calcium (Ca), and magnesium (Mg) with a high accuracy using data from the Moon Mineralogy Mapper (M3) on the Chandrayaan-1 mission. By analyzing the NIR reflectance of the lunar regolith, they develop a robust combination of spectral parameters that account for soil maturity effects, calibrating these parameters against elemental abundances from the Lunar Prospector Gamma Ray Spectrometer (LP GRS) and the Kaguya GRS (KGRS). Their results yield global elemental maps that align well with previous data, providing critical insights into lunar geology and evolution. This study relates to our findings by reinforcing the importance of hyperspectral imaging and spectral data analysis in understanding lunar mineralogy. Both our research and that of \cite{Bhatt2019} utilize advanced techniques to extract meaningful mineralogical information from spectral data, ultimately enhancing our understanding of the Moon's surface composition. Their emphasis on minimizing artifacts from space weathering aligns with our goal of accurately identifying olivine and pyroxene in volcanic analogs. Additionally, the elemental maps generated by \cite{Bhatt2019} could complement our mineral composition analyses, offering a more comprehensive picture of the processes that shaped both lunar and terrestrial volcanic landscapes. Additionally, our results align with insights from \cite{Thoresen2024}, who employed machine learning techniques to cluster lunar mineral data from  the Moon Mineral Mapper
(M3)\cite{SURKOV2020} and in \cite{Bhatt2019}, highlighting the effectiveness of K-Means in spectral classification. Their focus on global mineral distribution complements our localized analysis, providing a broader context for understanding lunar analogs. \cite{Zhang2021} further contribute to this discourse by exploring the spectral characteristics of iron-bearing minerals and offering methods for distinguishing between olivine and pyroxene based on diagnostic absorption features. Together, these studies advance our comprehension of mineral distribution and formation mechanisms in both terrestrial and lunar volcanic landscapes, enhancing our understanding of the similarities between volcanic processes on Earth and the Moon.

Future research should refine mineral definitions and test the methodology across a wider range of spectra to improve its applicability to lunar geology. Our forthcoming second paper will directly compare lunar observations and samples, providing insights into the relationship between terrestrial and lunar mineralogy.

\section{Discussion}\label{sec:discussion}
This study conducted a thorough investigation into the mineralogical composition of volcanic samples from Vulcano, a volcanically active island in the Aeolian Archipelago, north of Sicily, Italy, with a particular focus on the unique spectral characteristics of each sample. By systematically selecting representative sections and dividing them into nine distinct regions of interest, we effectively computed average spectral data that captured the mineralogical diversity present in the samples. This structured approach provided a nuanced understanding of variations in mineral compositions across different areas of the volcanic material. The results underscore the efficacy of various clustering algorithms applied to the analysis of olivine and pyroxene compositions. Using Principal Component Analysis (PCA) as a preliminary step, we successfully reduced the dimensionality of complex spectral data obtained from volcanic samples, which are terrestrial analogs to lunar minerals. This reduction facilitated clearer identification of patterns in mineral compositions, as illustrated in Figure \ref{fig:PCA}. The clustering algorithms assessed included K-Means, Hierarchical Clustering, Gaussian Mixture Model (GMM), and Spectral Clustering. The performance of these algorithms varied across different regions, as indicated by silhouette scores ranging from 0 to 1, with higher scores suggesting better-defined clusters.

K-Means clustering effectively partitions data into distinct clusters by minimizing variance within each cluster, and it is favoured for its simplicity and efficiency with large datasets. For instance, in Region 8, K-Means achieved a silhouette score of 0.47, the highest among all methods, indicating excellent cluster separation. Hierarchical clustering, which builds a hierarchy of clusters through either a bottom-up or top-down approach, provided a visual representation of data structure in a dendrogram format. In Region 7, this method achieved a silhouette score of 0.45, reflecting well-separated clusters. The Gaussian Mixture Model (GMM), a probabilistic model assuming data points arise from a mixture of several Gaussian distributions, showed flexibility in clustering. Notably, in Region 8, GMM matched K-Means with a silhouette score of 0.47, indicating similar clustering performance. Finally, Spectral Clustering, which utilizes eigenvalues of a similarity matrix for dimensionality reduction before clustering, achieved a score of 0.44 in Region 7, demonstrating moderate cluster separation.

Consequently, this study demonstrated the efficacy of hyperspectral imaging (400–1000 nm) combined with unsupervised clustering algorithms to analyze volcanic samples from Vulcano, Italy, as analogs to lunar materials. K-Means and Hierarchical Clustering outperformed GMM and Spectral Clustering in identifying olivine and pyroxene compositions, with K-Means achieving the lowest RMSE considering the scatter error. Our findings suggest a predominance of olivine, consistent with lunar volcanic processes, enhancing our understanding of mineral distribution. Future work should refine mineral definitions and test the methodology against diverse spectra to strengthen its applicability to lunar geology. Additionally, our second paper in this series will provide a direct comparison with lunar observations and samples, contributing valuable insights into the relationship between terrestrial and lunar mineralogy.

\section{Conclusion}\label{sec:conc}
Overall, the analysis indicates that all methods performed well with pure olivine compositions, achieving similarity percentages above 90\% in several clusters. However, when examining mixed compositions of olivine and pyroxene, performance variability became evident. K-Means and Hierarchical Clustering displayed better adaptability to mixed data, while GMM struggled with lower similarity scores in certain instances. These findings reveal significant details about the mineralogical composition of volcanic samples from the region, particularly in identifying lunar minerals like pyroxene and olivine \cite{CLENET2011}. The strong correlations established between spectral data and these minerals enhance our understanding of the region's volcanic activity and its geological evolution over time, with implications for lunar geology \cite{Dhingra2018}.
Our findings indicate a predominance of olivine in the Vulcano region, likely linked to volcanic processes akin to those on the Moon, as highlighted by \cite{Head1976}. \cite{SURKOV2020} map ilmenite abundance using Chandrayaan-1 M3 data, offering insights into lunar mineralogy that complement our focus on olivine and pyroxene. Both studies emphasize the importance of specific minerals in understanding the Moon's geological history. Likewise, \cite{Bhatt2019} quantitatively estimate elemental concentrations on the lunar surface using hyperspectral NIR imaging, reinforcing the significance of spectral data for mineral identification. Their global elemental maps enhance our understanding of lunar composition and align with our findings on volcanic analogs.
In terms of mineral identification, the analysis predominantly recognized olivine in the samples, though some methods also detected pyroxene. Olivine-rich compositions are common in ancient lava flows and impact melt rocks, as noted in previous studies \cite{McSween2006,Staid2011}. This is relevant to lunar geology, where olivine is a significant component of the Moon's crust, especially in areas shaped by ancient volcanic activity and meteorite impacts. Together, these studies deepen our understanding of mineral distribution and formation mechanisms in both terrestrial and lunar landscapes, underscoring the similarities in volcanic processes on Earth and the Moon.

The findings also highlight that the choice of clustering algorithm significantly impacts performance metrics; specifically, K-Means is optimal for certain clusters, while Hierarchical and Spectral methods excel in others. In the literature, Support Vector Machines (SVM) have demonstrated superiority for extraterrestrial bodies \cite{Sabat2020}, particularly when combined with Joint Mutual Information Maximization (JMIM) for higher accuracy \cite{Bhatt2022}. While our unsupervised approach suits exploratory analysis, future work could integrate SVM and JMIM to validate clustering results against labeled data, enhancing real-world applicability.

\vspace{6pt} 

\authorcontributions{Conceptualization: F.F.H.; Methodology:F.F.H. and M.R.; Software: F.F.H and M.R.; Validation: M.R., B.F., and F.J.V.; Formal Analysis: F.F.H; Investigation: F.F.H.; Resources: M.R. and P.S.; Data Curation: M.R. and P.S. and M.J.A.D; Writing---Original Draft Preparation: F.F.H and M.R.; Writing---Review and Editing: M.R. and M.J.A.D; Visualization: F.F.H.; Supervision: M.R., F.J.V., and B.F.; Project Administration: F.F.H. and M.R.; Funding Acquisition: F.F.H. and B.F.; E.C. and C.Z. assisted in coordinating the sample collection. All authors have read and agreed to the published version of the manuscript.}

\dataavailability{The data underlying this article will be shared on reasonable request to the corresponding author.}

\funding{P.S. and M.J.A.D. acknowledge support from the  Dutch National Science Agenda under project NWA.1418.22.022 which is made possible by financial support of the Dutch Research Council (NWO).}

\acknowledgments{F.F and M.R would like to express heartfelt appreciation to EuroSpaceHub and LUNEX EuroMoonMars Earth Space Innovation for their generous funding and unwavering support.}

\conflictsofinterest{The authors declare no conflicts of interest.}




\reftitle{References}


\bibliography{bibliography.bib}


\PublishersNote{}
\end{document}